# Proximity effect in granular superconductor-normal metal structures

I. Sternfeld[1], V. Shelukhin[1], A. Tsukernik[2], M. Karpovski[1], A. Gerber[1] and A. Palevski[1]

[1]School of Physics and Astronomy, Tel Aviv University, Tel Aviv 69978, Israel

[2]University Research Institute for Nanoscience and Nanotechnology, Tel Aviv University, Tel Aviv 69978, Israel

We fabricated three-dimensional disordered Pb-Cu granular structures, with various metal compositions. The typical grain size of both metals is smaller than the superconductor and normal metal coherence lengths, thus satisfying the Cooper limit. The critical temperature of the samples was measured and compared with the critical temperature of bilayers. We show how the proximity effect theories, developed for bilayers, can be modified for random mixtures and we demonstrate that our experimental data fit well the de Gennes weak coupling limit theory in the Cooper limit. Our results indicate that, in granular structures, the Cooper limit can be satisfied over a wide range of concentrations.

## 1. INTRODUCTION

Superconductivity in granular mixture films was extensively studied, both experimentally and theoretically, in the framework of the percolation theory[1,2]. It was demonstrated that for a random mixture of granular superconducting and insulating materials the superconductor-insulator transition occurs at the percolation threshold $P_C$, and the critical current of such systems vanishes at $P_C$[1].

When grains of the superconducting material are large enough, their intra-granular critical temperature is not affected by the proximity of the insulator component and remains close to the bulk value both above[3] and below[4] the percolation threshold. The macroscopic transition temperature can be reduced due to a gradual suppression of the inter-granular Josephson coupling across the metal – insulator transition[5].

Other studies considered the percolation and electronic properties of YBaCuO-normal metal compounds[6]. It was shown that in such samples the transition temperature is basically not affected as long as there is an infinite percolating cluster of superconducting component. It is important to emphasize that the size of the YbaCuO grains was larger than $\xi$. However the random mixture with opposite relationship namely with $\xi > d$ has never been investigated.

Random mixture films composed of a type 1 superconductor and a normal metal have never been investigated. For such systems one would expect that the intrinsic intra-granular superconductor transition temperature would be strongly affected by the presence of the normal metal due to the proximity effect. Thus far the proximity effect was investigated in bilayers and multilayers of superconducting (S) and normal (N) materials, where the variation of $T_C$ was studied as a function of the normal metal to superconductor thickness ratio $t_N/t_S$. Theoretical expressions for both weakly[7] [8] and strongly[9] coupled superconductors were derived for the dependence of $T_C$ on $t_N/t_S$ in the so-called Cooper limit, where the typical thickness of both the N and S regions are smaller than the relevant coherence length, $\xi_{N,S}$. A number of experiments were performed in order to determine $T_C$ for different superconducting and normal metal bilayers, such as Pb-Cu[10,11,12,13], Pb-Ag[14,15] and Nb-Cu [16]. The published experimental results for $T_C$ in Pb-Cu[10,11,12] bilayers are higher than those predicted by the theory of weak-coupling superconductors[8] and could be better fitted to the theoretical expression for the strong coupling case[9], or by modifying the weak-coupling formula by introducing a finite transparency of the barrier at the interface[13].

Recent investigation of the proximity effect in thin multilayered structure of Pb-Ag[14] also favored the theory of the proximity effect as derived for the strongly-coupled superconductor, however the agreement was poor for ultrathin films. There was also an attempt to fit the observed variation of $T_C$ in Pb-Ag films to the theory of the weak-coupling limit with a fitting parameter, taking into account the degradation of $T_C$ in the Pb grains due to the finite size effects[15].

In this paper we show how the theories developed for superconductor-normal metal bilayers can be modified for random mixtures and we demonstrate that our experimental data for Pb-Cu samples fit the weak-coupling limit theory very well. The appropriate parameter influencing $T_C$ in random mixtures (with typical grain size d<< $\xi_{N,S}$) is the ratio of volume concentrations- $P_N/P_S=P_N/(1-P_N)$ rather than $t_N/t_S$. We argue that for fine grains of both superconductors and normal metals, the Cooper limit is satisfied over a wide range of volume concentrations and, in contrast to a bilayer system, is not only limited to very thin films.

## 2. SAMPLE PREPARATION

Four terminal geometry samples were prepared with a standard photolithography technique. Gold ohmic contacts were evaporated using an e-gun. Then the substrate was transferred to the sputtering chamber and pumped to below $10^{-6}$ Torr. Samples were prepared by co-sputtering of Pb and Cu on the substrate. The sputtering target was composed of 16 pieces each made of either Pb or Cu. Such an arrangement allowed us to change the volume concentrations in the samples by varying the ratio of Pb and Cu pieces in the target. Volume concentrations could also be altered by the sputtering power. The total thickness of the Pb-Cu mixture films was about 100nm. A 50nm protective layer of Ge was evaporated *in situ* on top of the films. EDS X-ray analysis was performed for each sample in order to determine the chemical composition and ratio of volume concentration, $P_N/P_S = P_N/(1-P_N)$.

Transport measurements were preformed in an $He^4$ cryostat and Dilution refrigerator. Standard ac lock-in technique was used in these measurements.

## 3. EXPERIMENTAL DATA

Figure 1 shows TEM micrographs of a typical Pb-Cu film. As is evident from this figure, our films are composed of randomly spread Pb and Cu grains. The Pb grains are somewhat larger than the Cu grains, yet all grains are smaller than coherence lengths ($\xi_{N,S} \geq 30nm$) thus satisfying the Cooper limit. Bright field measurements that were taken indicate that samples are homogeneous.

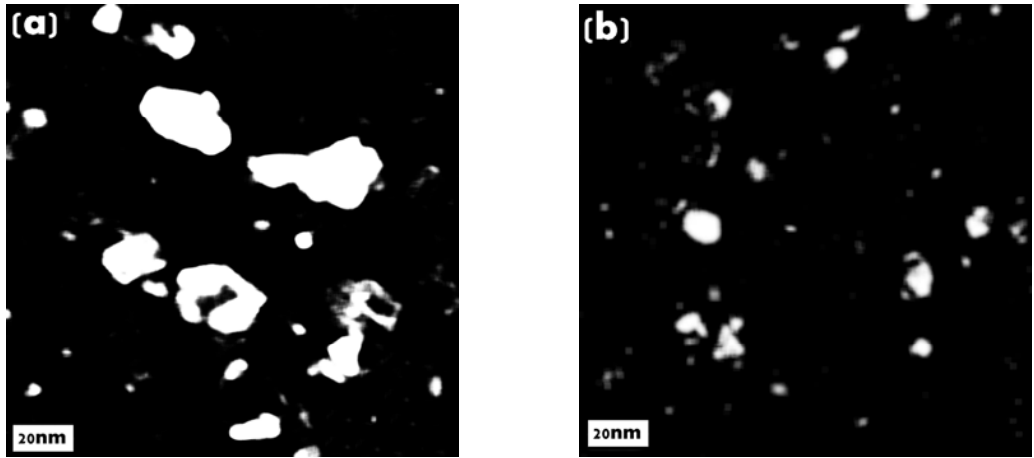

**FIG. 1. TEM micrographs of a typical granular Pb-Cu film. (a) Pb Dark field micrograph. (b) Cu Dark field micrograph.**

X-ray diffraction (XRD) was used for material structure characterization. XRD patterns were collected with CuKa radiation on the theta - theta powder diffractometer "Scintag", equipped with a liquid nitrogen cooled Ge solid-state detector. Unaltered patterns of immiscible Pb and Cu elements were superimposed on the XRD diffractogram, indicating that the grains of different samples are of equal purity.

Figure 2 shows the superconducting transition curves for four samples with different normal metal concentrations. The resistance values are normalized to the respective values at $T_{C0}=7.2K$. The superconductor transitions are rather sharp and quite smooth indicating the homogeneity of the samples. In contrast to the superconductor-insulator mixtures, $T_C$ is strongly affected by the variation of the concentration of the superconducting component. Similar to the behavior observed in bilayers, $T_C$ is strongly reduced from its bulk value, $T_{C0}$, even at concentrations with an infinite cluster of Pb percolating from one side of the sample to the other.

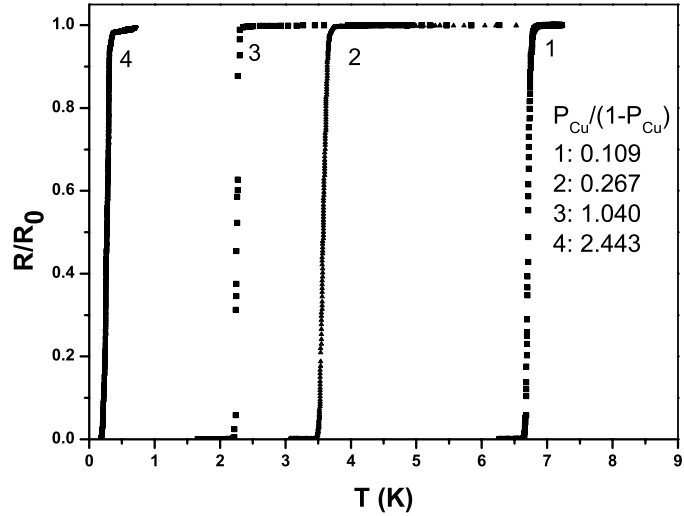

**FIG. 2. Normalized resistance $R/R_0$ as a function of temperature, for typical granular Pb-Cu samples with different volume concentrations. $R_0$ is the resistance at 7.2 K.**

As will become clear from our discussion below the appropriate parameter describing the degradation of $T_C$ in the Cooper limit for our granular mixture films is the relative volume concentrations $P_N/(1-P_N)$, rather than $t_N/t_S$ used for bilayers. The variation of $T_C$ versus $P_N/(1-P_N)$ is shown in figure 3. $T_c$ has been defined as the temperature at which resistance reaches either 1% (squares) or 50% (triangles) of the normal state values. Figure 3. indicates once more that the transitions are rather sharp and that the precise way we define our critical temperature is not important. However, in the appendix we shall show that in such granular films it is appropriate to define the critical temperature as the temperature at which the samples attain zero resistance.

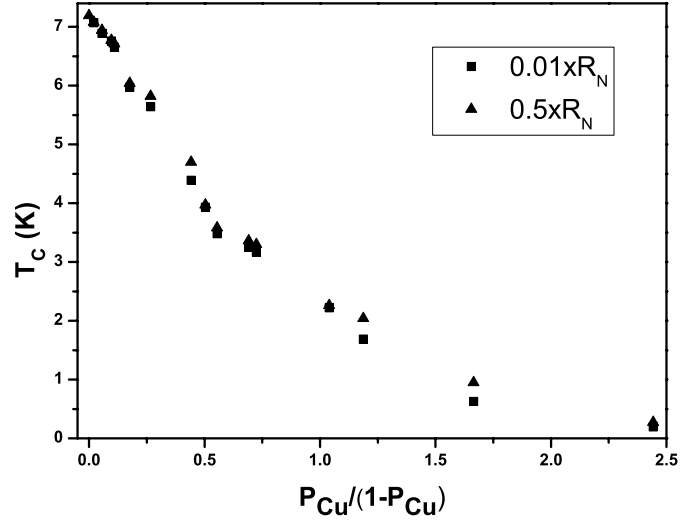

FIG. 3: Critical temperature $T_C$ versus relative volume concentration $P_{Cu}/(1-P_{Cu})$. The squares and triangles represent critical temperatures extracted from the temperatures at which the resistance attained 1% and 50% of the normal state resistance respectively.

## 4. ANALYSIS OF THE RESALTS AND DISCUSSIONS

It is well known[1,2] that many properties of random mixtures are successfully described by percolation theory. Experimentally, such systems were usually investigated for normal metal-insulator or superconductor-insulator mixtures and critical behavior was observed near the percolation threshold, $f_C$, and the appropriate critical exponents for the critical current, critical magnetic field, etc. were determined. It should be mentioned that the critical temperature of the superconductor-insulator film was not affected considerably by the proximity to $f_C$ and remained quite close to the bulk value of the superconductor as long as there existed an infinite superconductor cluster at f>f_C. One of the most important parameters in the percolation problem is the correlation length, $\xi_P$, which sets the homogeneity scale of the percolation network. On the length scale L above $\xi_P$, all the physical properties are assumed to be scale independent and vary only with the concentration, P, of the insulator or superconductor. For the superconductor proximity effect, the percolation of the infinite cluster is irrelevant since the electrons are no longer confined to the superconductor clusters but rather diffuse into the normal metal via Andreev reflections at the superconductor normal metal interface. We conject that the proximity effect in the random mixture films could be described as being similar to bilayers in the Cooper limit, provided that in all regions of size $\xi_N$ and $\xi_S$ there is a finite concentration of normal metal P (0<P<1). The latter requirement is analogous to the Cooper limit criterion for bilayers (see further discussion in the appendix).

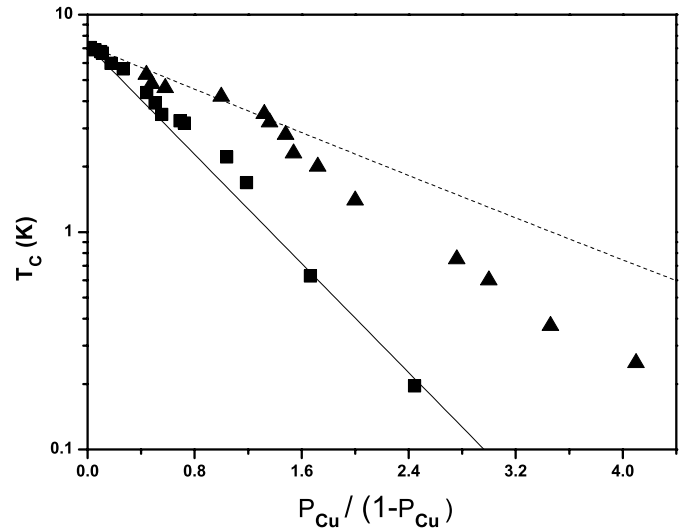

Fig. 4: Critical temperature versus relative volume concentration. The squares and triangles represent our and Z. Ovadyahu *et al.*[11] points respectively. The solid and

**dashed lines represent the weak[8] and strong[9] coupling limits respectively. In the theoretical expressions and bilayer data[11] the ratio $t_N/t_S$ was replaced by $P_N/(1-P_N)$.**

First we would like to compare our experimental data with the published data for Pb-Cu bilayers, by Z. Ovadyahu e*t al.*[11]. By courtesy of the authors we reproduce this data in Fig. 4 together with our data. One can clearly see that our experimental points in the entire range of volume concentrations lie well below the values obtained for the bilayer system. One of the reasons could be that the published data for the bilayers does not satisfy the Cooper limit criterion, however it is very unlikely since the authors claimed that the thickness used for Cu and Pb were smaller by at least a factor of three than the relevant coherence lengths. Another reason could be that the interface between the Cu and Pb obtained in a sequential deposition of the materials was partially oxidized and therefore had lower transparency than the interfaces between the grains obtained during the co-sputtering of our samples. Indeed, the low transparency interface between the superconductor and normal metal was considered in several publications[13,17,18] and it was demonstrated that the suppression of the critical temperature was weaker for less transparent interfaces.

In order to compare our data with the theories for the bilayers we plotted in Fig. 4 two theoretical curves [8,9] for different limits of superconducting electron phonon coupling constants- $\lambda$ :

$$\lambda > 1: \quad \ln(1.45 * T_c) = \frac{\lambda_S \ln \Theta_S + \lambda_N \ln \Theta_N \alpha^{P_N/P_S}}{\lambda_S + \lambda_N \alpha^{P_N/P_S}} - \frac{\lambda_S + 1 + (\lambda_N + 1)\alpha^{P_N/P_S}}{\lambda_S - \mu + (\lambda_N - \mu)\alpha^{P_N/P_S}} \quad (1)$$

$$\lambda << 1: \quad \ln \frac{T_c}{1.14\Theta_S} = -\frac{1 + \alpha^{P_N/P_S}}{(N_{S(0)}V)} \quad , \quad \alpha = N_N/N_S \quad (2)$$

Where $P_{N,S}$ are the normal metal and superconducting volume concentrations, λ is the coupling constant, μ is the (repulsive) Coulomb coupling constant, Θ the Debye temperature, $N_{(0)}$ the density of states at the Fermi level, V the attractive potential and $\alpha$ the ratio of density of states of the normal and superconductor materials.

Both theoretical curves are calculated for the following microscopic parameters [9,14]: $\Theta_{Pb}$=102 K, $\Theta_{Cu}$=343 K, $\mu$=0.13, $\lambda_{Cu}$=0.15, $\lambda_{Pb}$=1.013 and $N_{Pb(0)}V$=0.36. The last two parameters were obtained by posting $P_{Cu}/(1-P_{Cu})$=0 in both formulas and comparing with the bulk value critical temperature. Using the relation $N \propto m_e^* n^{1/3}$, where $m_e^*$ is the electron effective mass and n is the free electrons density, we calculated $\alpha$ to be 0.52[19].

Although lead is assumed to be a strongly coupled superconductor, our data points only deviate slightly from the curve of the weakly coupled limit. To the best of our knowledge all the reported experimental results for Pb-Cu bilayers[10-12] favor the strong coupling expression and, more generally, no published experimental data for Pb and a normal metal follow the weak coupling expression closely[14], unless a fitting parameter is used[13,15].

## 5. CONCLUSIONS

In summary, we demonstrated that three dimensional granular mixture films are unique systems for investigating the superconducting proximity effect in a wide range of relative volume concentrations. We argue that in such systems the Cooper limit is satisfied without any limitation on the total thickness of the film, provided that grains of both constituents composing the mixture are really fine, namely $d_{S,N} \ll \xi_{S,N}$. Moreover, the high quality of the interface obtained during the co-sputtering guarantees ideal electronic transmission, which is usually assumed in the existing theoretical models. We would like to emphasize that our data is in reasonably good agreement (without any fitting parameters) with the expression derived for the weak coupling limit.

## 6. ACKNOWLEDGMENTS

We would like to thank K. Efetov, A. Volkov, G. Schön, A. Aharony, O. Entin-Wohlman, Y. Kantor, and G. Deutscher for fruitful discussions, Dr. Yuri Rosenberg from the Wolfson Applied Materials Research Center of Tel-Aviv University and Y. Feedman from the Department of Materials and Interfaces of the Weizman Institute of Science in Rehovot, Israel for their assistance.

This research was supported by a grant from the German-Israeli Foundation for Scientific Research and Development and the Israel Science Foundation, founded by the Israel Academy of Sciences and Humanities-centers of Excellence Program.

## APPENDIX

The purpose of this section is to show that the proper factor which describes the decrease of $T_C$ in our granular films is the relative volume concentrations $P_N/(1-P_N)$, rather than $t_N/t_S$ used for bilayers, to determine this factor in the more general case and to illustrate that $T_C$ should be defined as the temperature at which the film reaches zero resistance.

Let us divide our sample into cubic cells of linear dimension $\xi$ ($\xi = \xi_S \simeq \xi_N$). Since $\xi$ is finite, each cell will have a slightly different concentration of normal metal P. The fraction of cells having concentration P can be described by a normal distribution function $N_{<P>,\sigma}(P)$, where <P> is the samples average of normal metal concentration and $\sigma^2$ its variance which depends on $\xi/d$. Each cell will have its own critical temperature, $T_C$, determined by the local P of the cell, independent of the $T_C$ of the neighboring cells, because the correlations between the superconducting properties of each cell vanish rapidly over distances exceeding the coherence length $\xi$. This implies that the $T_C$ of a cell can be calculated using Eq. (1) or (2) where P/(1-P) is substituted instead of $t_N/t_S$.

At a given temperature T, the volume fraction of the sample that is in the superconducting state can be obtained from $f$ - the fraction of the superconducting cells:

$$f = \int_0^{P(T)} N_{<P>,\sigma}(P)dP \quad (3)$$

where the upper limit of integration is the concentration of normal metal for a cell that becomes superconducting at T. At a certain fraction, $f_C$ (percolation threshold), of superconducting cells the sample will be spanned with a superconducting cluster and its resistance will vanish. $f_C$ should not be confused with Pc since they describe different percolation networks- the first being a network of cubic cells with linear dimensions $\xi$ and the latter being a network of grains with typical dimensions d. Therefore the $T_C$ of the entire sample will be determined by a $P(T_C)$ such that $f = f_C$ namely:

$$f_C = \int_0^{P(T)} N_{<P>,\sigma}(P)dP \quad (4)$$

Actually the degradation of $T_C$ in figures 3 and 4 should have been plotted versus $P(T_C)/(1-P(T_C))$ and not versus $<P>/(1-<P>)$. Below, however, we demonstrate that, in our case the relative deviation of $P(T_C)$ from $<P>$ is negligible. Indeed, the integral in Eq. (4) can be estimated by the well known error function:

$$1 - 2f_c = erf\{\frac{<P> - P(T)}{\sqrt{2}\sigma}\} \quad (5)$$

Using Eq. (5) and known values of $f_C$ from percolation theory we estimate the relative shift of $P(T_C)$ for three different cases:

1. A two-dimensional sample namely $\xi \geq h$ (h is the total film thickness): the percolation theory predicts $f_C = 0.5$ and since $erf(0)=0$, we obtain $P(T_C)=<P>$.

2. A three-dimensional sample, namely $\xi << h$: the percolation theory predicts $f_C = 0.16$. Using published tables for the error function and substituting $\sigma$ in terms of $<P>$ and $\xi/d$ into Eq. (5) we get:

$$\frac{<P> - P_{(T_C)}}{<P>} \approx \left(\frac{d}{\xi}\right)^{3/2} \sqrt{\frac{1-<P>}{<P>}} \quad (6)$$

So, for $<P>$ smaller than 0.44, the relative shift of $P(T_C)$ exceeds 10%.

3. In our samples the ratio $h/\xi$ is smaller than 3. The percolation theory predicts that $f_C$ of such a sample is about 0.38[20] and in this case:

$$\frac{<P> - P_{(T_C)}}{<P>} \approx 0.3 \left(\frac{d}{\xi}\right)^{3/2} \sqrt{\frac{1-<P>}{<P>}} \quad (7)$$

The relative deviation of $P(T_C)$ from $<P>$ will exceed 10% only for samples with $<P>$ smaller than 0.07. Only two of our samples have $<P>$ smaller than 0.07 and for these samples we calculate the absolute shift and find it to be about 0.005 which is smaller than the accuracy of the concentration measurement.

We assumed here, for simplicity, that $\xi$ is constant. It is well known however, that $\xi_{(T)}$ varies as $\xi_{(T)} \sim 1/\sqrt{T}$ [8], hence the relative deviation of $P(T_C)$ from $<P>$ in Eq. 7 is reduced due to two reasons: first the numerical factor is reduced due to dimensionality cross over and eventually vanishes at $\xi_{(T)} \geq h$, second the ratio $d/\xi_{(T)}$ becomes smaller at lower temperatures.

We conclude that, for our films, the temperature in which the sample attains zero resistance can be approximated by formulae (1) and (2) derived for bilayers replacing $t_N/t_S$ with $<P>/(1-<P>)$. However, for three dimensional samples and somewhat larger grains, the fluctuations of the concentration could be important and therefore $<P>$ should be replaced with $P(T_C)$ as in Eq. (4) and (5).